\pdfoutput=1
\documentclass{itechreport}
\usepackage{graphicx}
\usepackage{algorithm}
\usepackage{algorithmic}
\usepackage{subfigure}
\begin{document}
\title{Storage Balancing in Self-organizing Multimedia Delivery Systems}
\author{Anita Sobe \and Wilfried Elmenreich \and Laszlo B\"osz\"ormenyi}
\itechnumber{TR/ITEC/01/2.13}
\maketitle
\begin{abstract}
Many of the current bio-inspired delivery networks set their focus on search, e.g., by using artificial ants. If the network size and, therefore, the search space gets too large, the users experience high delays until the requested content can be consumed. In previous work, we proposed different replication strategies to reduce the search space. In this report we further evaluate measures for storage load balancing, because peers are most likely limited in space. We periodically apply clean-ups if a certain storage level is reached. For our evaluations we combine the already introduced replication measures with least recently used (LRU), least frequently used (LFU) and a hormone-based clean-up. The goal is to elaborate a combination that leads to low delays while the replica utilization is high.
\end{abstract}
\section{Introduction}
The increased complexity of modern networks and the increasingly dynamic access patterns in multimedia consumption have led to new challenges for content delivery. Bio-inspired networks promise a robust and adaptive behavior to cope with this task~\cite{harsh:10}.
We rely on a bio-inspired delivery algorithm introduced in \cite{Sobe2010a} and its extension in \cite{Sobe2011}, providing the possibility of sharing small multimedia units during and after a social event. A multimedia unit can be either a photo or a short video sequence, e.g., generated at a live sports event by a visitor. With the model of small units two typical problems of delivery do not exist anymore: (1) The transport of large files, as typical for multimedia content; (2) Content consistency, since units do not change after creation. However, we assume dynamic access patterns. Users can "compose" their presentations consisting of a number of different units. E.g., a visitor of a sports event wants to see an overview of the highlights of the last 20 minutes presented as 4 units in parallel in a split-screen. Such access patterns allow flexibility, but also introduce complexity, because the typical video sequence as known from movies is not existing anymore.\\
Our delivery algorithm is capable of handling this complexity relying on simple local information. The algorithm is inspired by two existing bio-inspired approaches. The first approach is Ant Colony Optimization as discussed in \cite{Dressler2010}; a specific application for search in P2P networks was introduced as SemAnt~\cite{Michlmayr2007a}. The second approach is an artificial hormone-based agreement for task allocation introduced by Brinkschulte et al. in \cite{Brinkschulte2007}. \\

We adopted the keyword search from SemAnt and introduce one type of hormone by keyword. The hormone value expresses the current demand (the \emph{goodness}, as adopted from Brinkschulte et al.) for the corresponding keyword. SemAnt assumes the content distribution to be fixed, which keeps the search space very large. To reduce the search space we apply replication and exploit the unstructured overlay by letting the content travel towards higher levels of its corresponding hormone. This allows intermediate nodes to decide if the traveling unit is needed for future requests, thus predicting places to reduce search space and delivery costs. \\

Since peers are not likely to provide unlimited storage for other peers and in a dynamic system the popularity of units changes, we investigate different measures to efficiently balance the storage of the peers. We periodically apply different strategies if a certain storage level is reached. We compare LRU (Least Recently Used), LFU (Least Frequently Used) and a hormone-based clean-up. We show that the chosen clean-up mechanism has an impact on the delivery performance and that the system is still robust against peer churn. We performed simulations for evaluating the replication and clean-up mechanisms in scale-free and random overlay networks.\\
\section{Hormone Algorithm Description}
\begin{figure}
  \centering
  \includegraphics[scale=0.7]{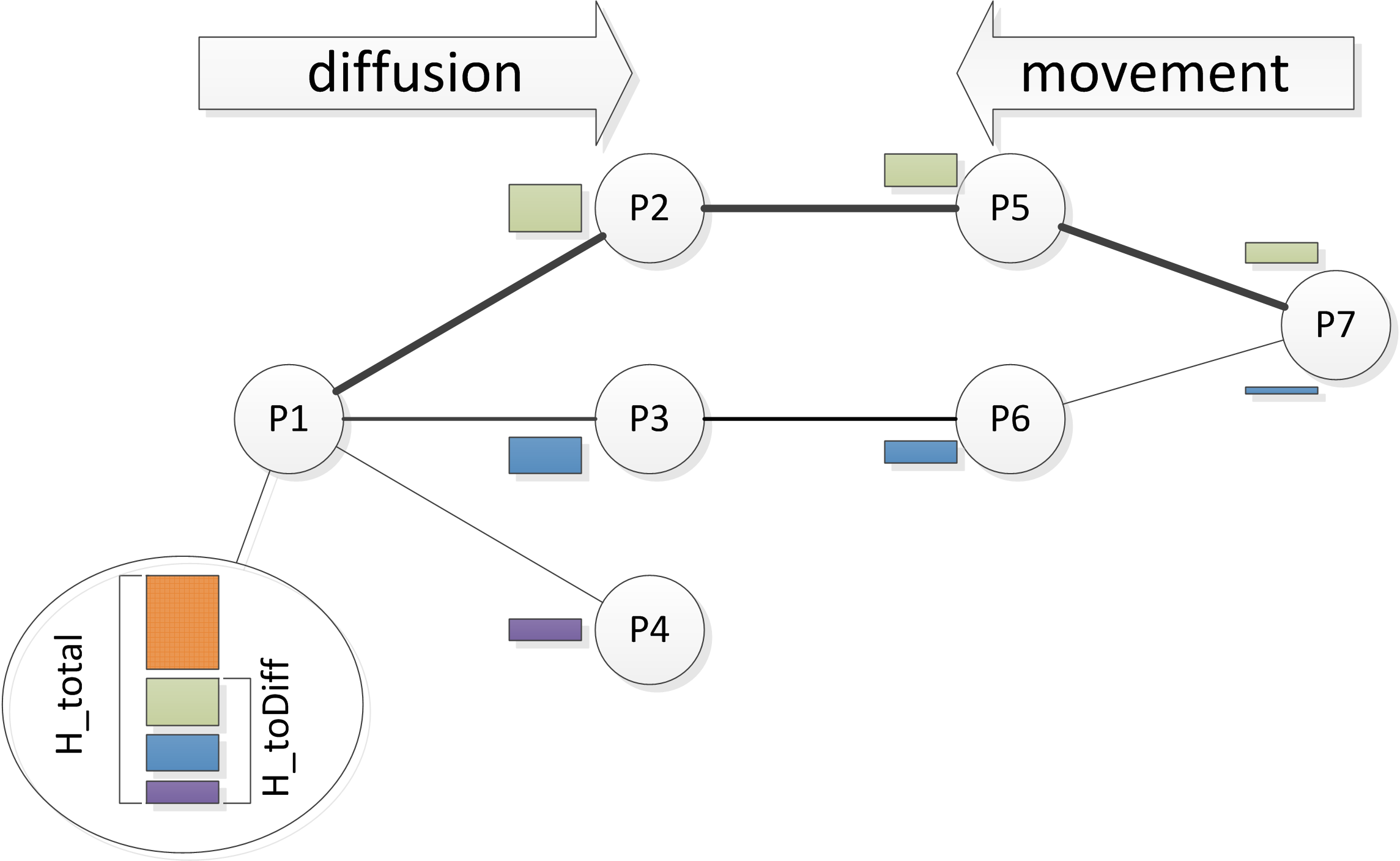}\\
  \caption{Interplay of diffusion and unit movement in hormone-based delivery}
	\label{fig:hormone}
\end{figure}
The hormone-based delivery approach introduced in~\cite{Sobe2010a} involves three components: hormone creation, hormone diffusion, and the behavior of units in presence of a corresponding hormone. The distributed, self-organizing nature of the approach allows to handle the complexity of requests and the search for units in the network with comparably simple decision algorithms based on local knowledge.\\
Upon request of a particular unit, a corresponding hormone is issued at the requesting node. The hormone is represented as a real value; the current value of a hormone is called \textit{hormone level}. The hormone is diffused via the neighboring nodes to the network, creating a hormone gradient towards the requesting node.\\

The diffusion of hormones depends on the network structure, e.g., a node that is only a few hops away from the requesting node will be flooded earlier with hormones and will get more hormones than a more distant node. \\
Figure~\ref{fig:hormone} depicts the diffusion of hormones from a requesting node $P_1$. A part from the total amount of deployed hormones is diffused to the neighboring nodes $P_2$, $P_3$, and $P_4$. Note that the amount of diffused hormone is spread unevenly, based on the condition of the network links to the respective nodes (a thicker line indicates a better/faster connection).\\

As long as a request is not fulfilled, the hormone level is raised periodically, thus increasing the \emph{pull} for a unit fulfilling the particular request.\\

A unit that corresponds to the given hormone will react by moving towards the network path with the steepest gradient, eventually arriving at the requesting node. In order to reduce attracting multiple copies of a unit, the diffusion of a hormone is stopped, if the respective unit is residing already on the same node.\\

The hormone-based delivery creates a feedback loop between network conditions. The network conditions influence hormone diffusion, the hormone diffusion influences unit movement, which in turn creates network traffic and changes the network conditions.\\

Multiple requests for different units lead to a set of different hormones being handled in parallel by the network. Requests for the same unit result in a superimposed hormone landscape for that unit. In this case, a unit might be attracted by two hormone trails. Without replication the unit has to move first to the first requester and afterwards to the second requester. Which requester gets the unit first depends on the strength of hormone reaching the unit. In order to avoid such detours, an intelligent replication mechanism has to take care of this issue.
\section{Parameter Settings}
\label{sec:params}

\begin{table}
\centering
\caption{Parameters to configure at system startup}
\label{tab:parameters}
\begin{tabular}{|c|l|}
  \hline
  ID & Explanation \\
  \hline
  $\eta_0$ & Hormone strength of a unit at new request \\
  $\eta$ & Increase of hormone after each time step by the requester\\
  $\alpha$ & Percentage of hormones to be forwarded to the neighbors \\
  $\epsilon$ & Hormone evaporation value \\
  $t$ & Minimum hormone strength\\
  $m$ & Minimum hormone difference to move unit\\
  \hline
\end{tabular}
\end{table}

The proposed algorithm is self-organizing and has only a few configuration parameters. In Table \ref{tab:parameters} the necessary parameters are shown. These parameters are dependent on each other. E.g., if $\eta_0$ and $\eta$ are low and the evaporation value $\epsilon$ is high, the movement of units can be limited. The more hormones are created and forwarded, the more hops the hormones can travel and therefore increase the search space. The difference to move a unit $m$ controls the mobility of units. If $m$ is high, the units need a higher hormone concentration to move to a neighbor, leading to a longer waiting time for the requester. In general, the parameter settings are essential for the algorithm to work. Therefore, we decided not to tweak the parameters manually, but optimize them by using a genetic algorithm. In this work a genetic algorithm such as in \cite{Elmenreich2007} is used.\\

Initially, the algorithm creates a random population. Then, it uses elite selection for building the next generations. The candidates are sorted according to their fitness and the best $x$ candidates are chosen. These candidates propagate to the next generation. To reach the same population size as the last generation, the rest of the slots are reserved for mutation, crossover and new candidates. For mutation and crossover random elite candidates are chosen. Finally, random new candidates are added to the population.\\

The fitness function targets client satisfaction, therefore, it optimizes the number of consumed units. The genetic algorithm is part of a simulator that also runs the artificial hormone-system. Depending on the number of generations the genetic algorithm starts the simulation of the artificial hormone-system and records the resulting parameters. The parameters have to be generated only once and can be used as input for the artificial hormone-system.

\section{Clean-up/Replacement Strategies}

We concluded in \cite{Sobe2011} that efficient replacement or clean-up has to be done in order to avoid blocking the transport of units.  A clean-up is triggered if a certain storage level is reached, which leads to a balanced storage load of the system.\\

However, if we want to guarantee that always at least one instance of the unit is available in the system, the decision of what unit to delete becomes more complicated. Therefore, the goal is to find an efficient strategy that neither influences the delay, but on the other side increases the utilization of replicas. We apply the basic principle, where none of the strategies deletes a unit if there is no copy on one of the neighbor nodes.\\

We compare three mechanisms least recently used (LRU), least frequently used (LFU) and hormone-based clean-up. LRU takes care of popularity changes of units. If a unit is not popular anymore its replicas can be removed. LFU targets units of low popularity. Hormone clean-up exploits local knowledge about hormone concentration. A unit is deleted if there are no hormones for it on the neighbors, thus there is currently no demand for it. So, units currently in delivery are not deleted.\\

We combine the results of the clean-up/replacement strategies with the replication strategies evaluated in \cite{Sobe2011}. In the following a short description of these strategies is provided:

\begin{itemize}
	\item \textbf{Owner Replication.} The content is replicated at the requester's node \cite{Lv2002}
	\item \textbf{Path Replication.} In a multi-hop network where content is not transported directly such as in Freenet \cite{Clarke2001}, it is possible to cache a replica of the content in transmission in each intermediate node. Since the intermediate nodes are acting as caches, the path replication is also called cache-based replication. It is assumed that intermediate nodes provide storage space for replicas even if they are not interested in the content. Path replication leads to a high number of unused replicas.
	\item \textbf{Path Adaptive Replication.} An alternative to path replication is to specify a node specific replication probability, where nodes decide ad-hoc if a file is replicated or not. The replication probability is dependent on the peer's resource status and optionally refers to the replication rate, too \cite{Yamamoto2005}.
	\item \textbf{Simple Hormone}. If a unit is requested by peers from opposite parts of the network, the unit has to move first to one requester and afterwards to the other requester. This can lead to long traveling paths, which can be avoided by replicating a unit if more than one neighbor holds hormones for it. Note that it is not possible to differentiate if hormones on the neighbor are created by different peers. Thus, it is possible that unnecessary replications are made.
	\item \textbf{Local Popularity}. Each node uses the local request history of the corresponding content to decide if it is likely to be requested again in the future. If the rank of a content is among the best 30 \% the corresponding unit is kept. So popular units are more likely to be replicated, but popularity information from neighbors is ignored. The communication effort is minimized.
	\item \textbf{Neighbor Popularity Ranking}. After collecting the popularity ranks for a content from the neighbors, the peer decides if it is worth to replicate the corresponding unit. The ranks are aggregated to a region rank (see \cite{Sobe2009b}), which is calculated as follows:
	\[R=\frac{1}{n} \sum_1^n{\ln(r_i)}\]
If the region rank $R$ is lower than a given threshold (e.g., the best 30 \% at all neighbors) the unit is replicated. $n$ represents the number of neighbors and $r$ is the rank of the specific unit at this neighbor. To reduce the impact of peak ranks (e.g., one unit is best ranked at two nodes, but worst ranked on the third node) the logarithm is used. The cooperation of neighbors is advantageous if their taste diverges.
	\item \textbf{Neighbor Hormone Ranking}. Analogue to the popularity ranking the units can also be ranked by their hormone values at the neighbors. The higher the hormone value for a unit on a neighbor is, the better is the unit's rank. The collected ranks can be aggregated as before and if the region rank is lower than a given threshold (e.g., the best 30 \%), the unit is replicated.	
\end{itemize}
\section{Simulation Settings}

We implemented a simulator that allows for the definition of network topology, storage behavior, client behavior, etc. and performs at certain time steps hormone management, request generation, unit movement and replacement. In the following the specific settings of the current evaluation are described.

\subsection{Network Topology}

We assume for small overlay networks of 50 nodes a connected Erd\H{o}s-R\'enyi random graph with a diameter of~6. For larger networks, e.g., with 1,000 nodes, we assume a scale-free network topology. To generate such a network the Eppstein Power Law Algorithm~\cite{Eppstein2002} is used. The algorithm gets as input a random graph and by repetitively removing and adding edges a power law distribution is reached. The network diameter of the scale-free graph is 13. The bandwidth was set to 100 Mbit/s. Note that further bandwidth scenarios and parameter studies are target of future work.

\subsection{Initial Storage}

Each node creates units until 30 \% of each node's storage is filled, where each node supports the system with 900 MB. At the beginning only one instance of each unit exists. We expect that in a scenario with 50 motivated persons, each person is contributing with equal probability. In a scenario with 1,000 visitors we expect that there are few highly motivated persons and a high number of less motivated persons. We further assume that each person is represented by one peer. We generate 5,000 units for the 50 peers scenario, and 15,000 units for the 1,000 peers scenario.\\

The average size of a unit is 2.6 MB, whereas the maximum size is 16 MB and the minimum size is 190 KB, with a playback bit rate of 1 Mbit/s. These unit sizes are the result of the third SOMA use case "The long night of research"\footnote{SOMA Web-page http://soma.lakeside-labs.com}.

\subsection{Request Generation}
We provide keyword search, where a number of units are supposed to match the same keyword. The number of units per keyword follows a Zipf-like distribution.\\

One request consists of a number of keywords. The request is fulfilled if for each keyword at least one unit is stored on the node. We further implemented a taste change, i.e., if a user likes the content just watched, her taste for future requests might be similar to the currently watched unit. \\

In this paper we do not consider any order of the units, thus, if a requested unit arrives, it is presented to the user. Sequential and parallel dependencies have been handled in~\cite{Sobe2010a}. \\

We further introduce a deadline for each unit, until which it has to be delivered. The deadline is dependent on the size, the link bandwidth and the maximum number of hops a unit can travel $maxhops$. If a deadline is missed, no further hormones for that unit are created to stop attracting content. \\

A request is considered as failed if none of the requested units could fulfill their deadline. A user can only submit one request at a time. If this request is fulfilled or failed, a new request will be generated.

\subsection{Simulation Parameters}

We used the genetic algorithm as described in Section \ref{sec:params}. We change the fitness function to maximize the number of successful requests (i.e., consumed units within the deadline). The optimized parameter set is used for both random and scale-free network scenarios.\\

In Table \ref{tab:pars} the resulting parameters are shown. On creation the hormone value is high, leading to a wider travel range. The diffusion of 45~\% of the hormones supports longer travel distances, too. The evaporation rate is in comparison to the creation number rather low, which means the hormones last for some time. The evaporation hormones are a fixed value subtracted from the current hormone level. The migration threshold describes the minimum hormone difference between two nodes to make a unit move. In this case the difference is very low in comparison to the creation amount of hormones. This leads to a very dynamic behavior of the units. The clean-up is triggered by a node if its current storage level exceeds 60~\%.
\begin{table}%
\centering
\begin{tabular}{|l|c|}
$\eta_0$ & 3.95\\
$\eta$ & 4.39\\
$\alpha$ & 0.45\\
$\epsilon$ & 0.16\\
$m$ & 0.23\\
$c$ & 60 \% \\
$t$ & 0.23 \\
$maxhops$ & 10
\end{tabular}
\caption{Parameter settings}
\label{tab:pars}
\end{table}

\subsection{Metrics}

We want to evaluate the request fulfillment on the one hand and the utilization of replicas on the other hand. The fulfillment of requests is represented by the \textit{delay}. The delay is measured from the request time of a unit until the arrival of that unit on the node. A delay of $0$\~s means that the unit was already on the node. The delay is presented as cumulative distribution function (CDF) over the simulation time. The \textit{deadline missed rate} represents the rate of units (not requests), for which the deadline is missed. If a unit missed its deadline, the delay is calculated as deadline minus request time (max. delay). The \textit{request failed rate} indicates requests from which all units missed their deadline.\\

A unit is presented for some time, and we measure the rate of units that currently started with presentation. The more unit presentations started in comparison to the number of their replicas, the better the \textit{unit utilization}. The utilization and the request failed rate will be depicted as box plot with 1.5 interquartile range whiskers. \\

\section{Evaluation}

\subsection{Scenario 1: 50 nodes network}
\begin{figure}%
\centering
\includegraphics[scale=0.4]{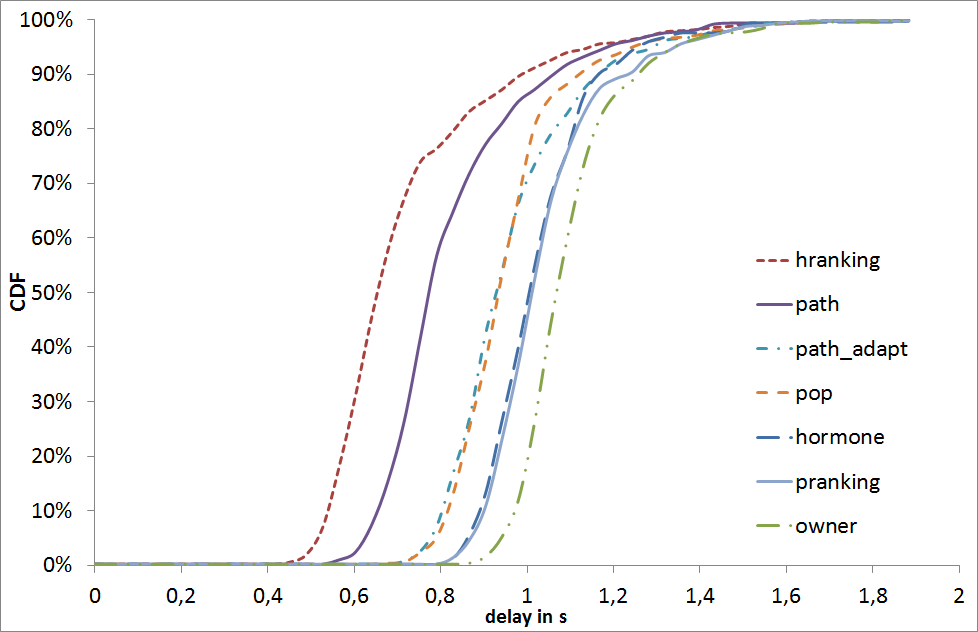}%
\caption{Delay Comparison of replication without clean-up \cite{Sobe2011}}%
\label{fig:nocleanup}%
\end{figure}
\begin{figure}%
\centering
\subfigure[Delay distribution of Hormone clean-up]{\includegraphics[scale=0.3]{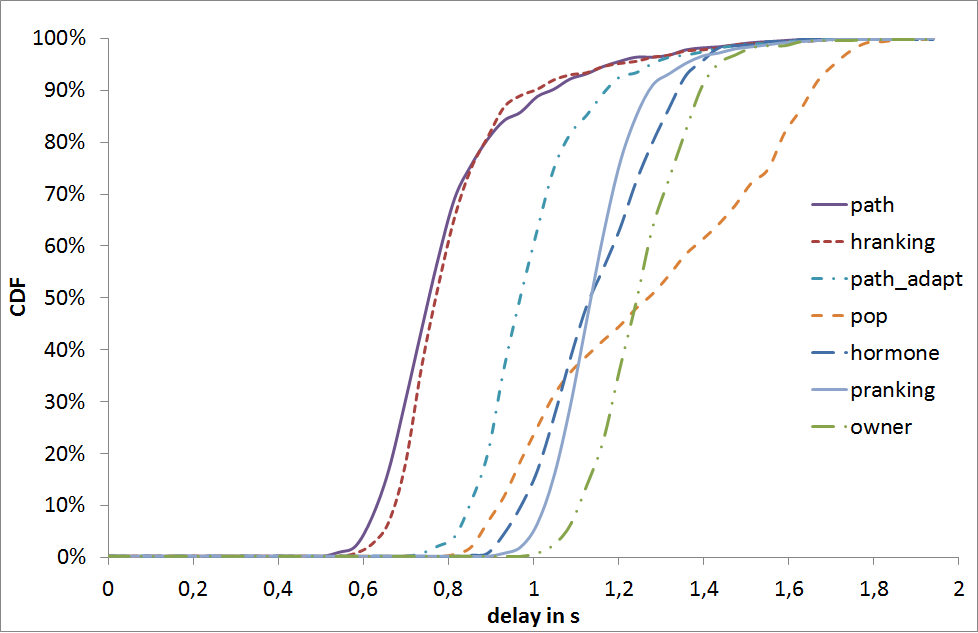} \label{fig:hcleanup}}
\subfigure[Delay distribution of LRU clean-up]{\includegraphics[scale=0.3]{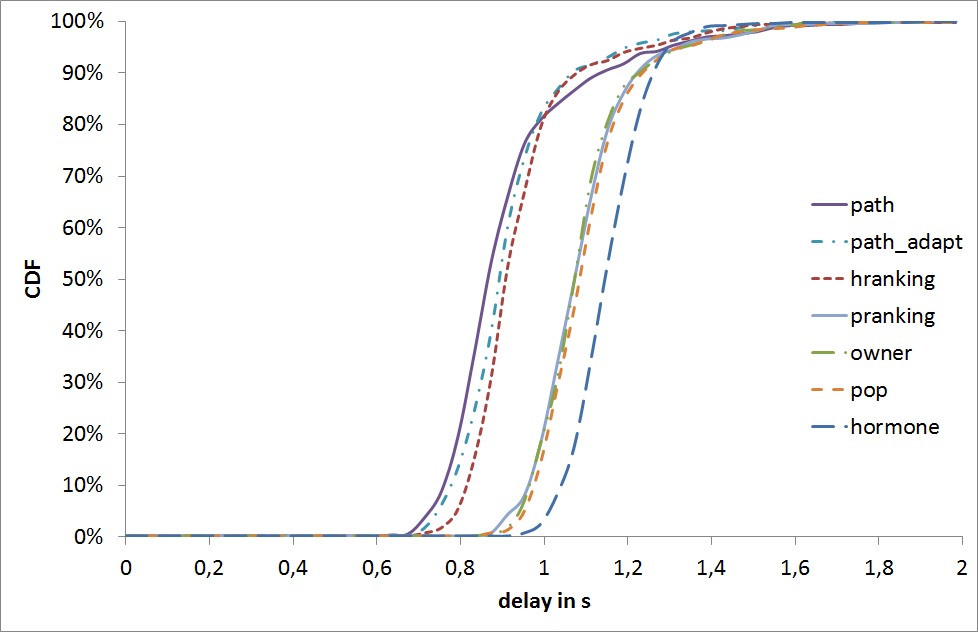} \label{fig:lrucleanup}}
\subfigure[Delay distribution of LFU clean-up]{\includegraphics[scale=0.3]{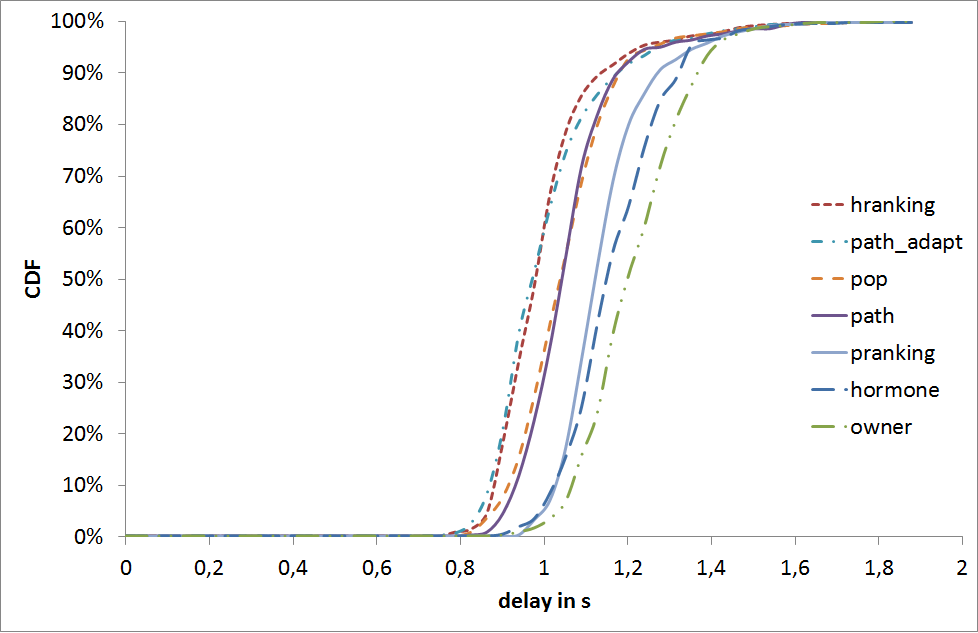} \label{fig:lfucleanup}}
\caption{Delay Comparison of the different clean-up mechanisms}
\label{fig:delay}%
\end{figure}
\begin{figure}%
\centering
\includegraphics[scale=0.4]{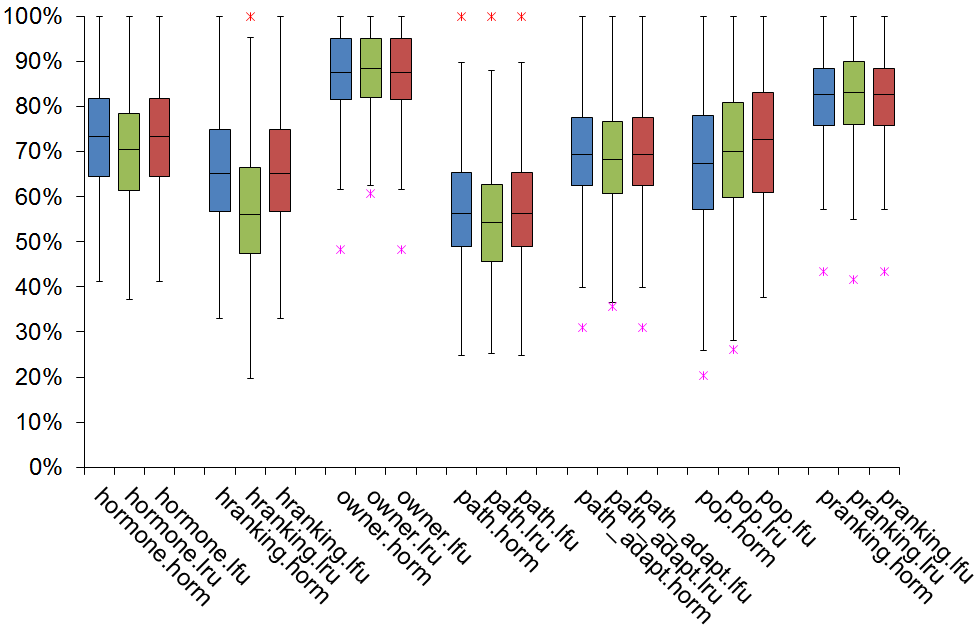}%
\caption{Hormone, LRU and LFU utilization comparison}%
\label{fig:utilization}%
\end{figure}
\begin{figure}%
\centering
\includegraphics[scale=0.4]{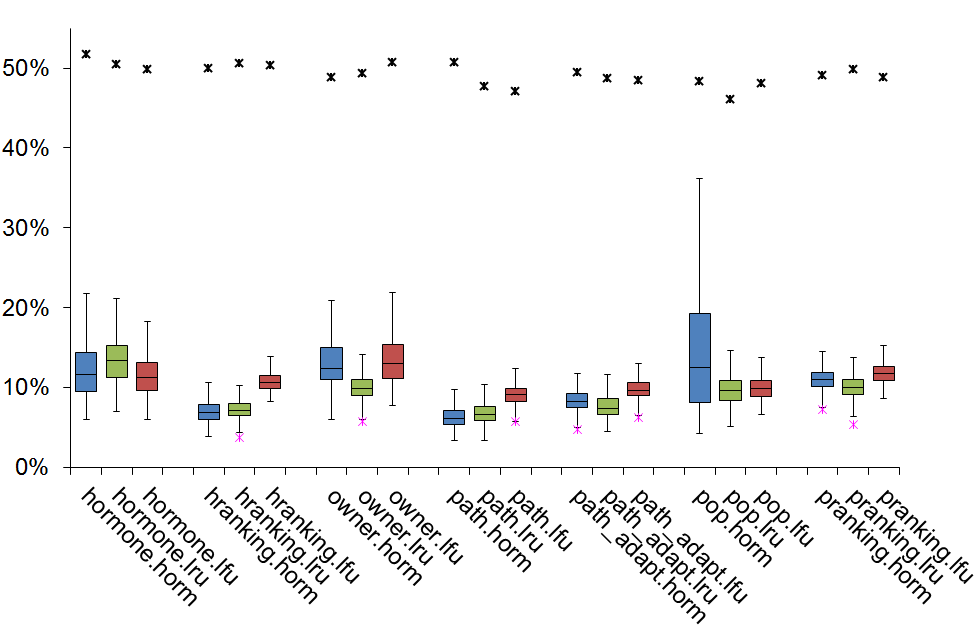}%
\caption{Hormone, LRU and LFU request failed rate comparison}%
\label{fig:failedReq}%
\end{figure}

In this section we concentrate on the evaluation of clean-up mechanisms if applied to the replication strategies discussed above. In Figure \ref{fig:nocleanup} we show again the delay distribution of the replication strategies without clean-upIn comparison to the pure replication shown in Figure \ref{fig:nocleanup} in Figures \ref{fig:hcleanup}, \ref{fig:lrucleanup} and \ref{fig:lfucleanup} hormone clean-up, LRU and LFU are applied. One can see immediately that some combinations of clean-up and replication are positive, whereas some combinations lead to performance drops. The most conspicuous example is the combination of local popularity replication and hormone clean-up as shown in Figure \ref{fig:hcleanup}. The reason for this is that the hormone clean-up only considers current demands for a unit and therefore does not fit the replication mechanism.\\

It is further interesting to see that path replication in combination with hormone clean-up results in a lower delay than hormone ranking. Hormone ranking initially places units more efficiently, leading to less replicas than path replication would create. The downside of hormone ranking is that popular nodes evolve, which are filled first. This leads to blocked transport paths, which in turn increases the delay.  \\

The LRU clean-up depicted in Figure \ref{fig:lrucleanup} has the most positive impact on path-random replication since the delay is stable in comparison to the best effort scenario, while the other replication mechanisms experience higher delay. \\

LFU (Figure \ref{fig:lfucleanup}) leads to a high number of wrong clean-up decisions, because the delay increases remarkably for all replication mechanisms. Especially the hormone replication mechanism suffers from wrong decisions; the curve gets flatter, indicating that the delay jitter increases. The local popularity mechanism experiences positive impact, since LFU also prioritizes popular units.\\

The utilization measures the improvement in storage efficiency of the single replication mechanisms (see Figure \ref{fig:utilization}). All clean-up strategies lead to an increase of utilization. Hormone replication does not take advantage of a clean-up, because it already creates before only a low number of replicas. Hormone ranking has the worst utilization if combined with LRU, but has even in that case a higher utilization than path replication. Local popularity replication results in a high variance, which also shows that some improvement is needed. \\

A high utilization does not necessarily indicate that the replication mechanism is best suited for the delivery. A high utilization can also be reached if deadlines are missed and therefore the delivery did not take place, which leads to a lower number of replicas. Therefore, it is important to aggregate the information of delay, utilization and the requests failed statistics.\\

In Figure \ref{fig:failedReq} one can see immediately the low performance of local popularity replication if combined with hormone clean-up. The lowest variance is reached by combining it with LFU. The other replication strategies do not profit from the combination with LFU, the request failed rate is the highest. Most replication strategies work best if combined with LRU, except hormone ranking. \\

The decision of what clean-up strategy to choose should be made by combining the results of delay, utilization and request failed rate.

\subsection{Impact of Peer Churn}
\begin{figure}%
\centering
\includegraphics[scale=0.35]{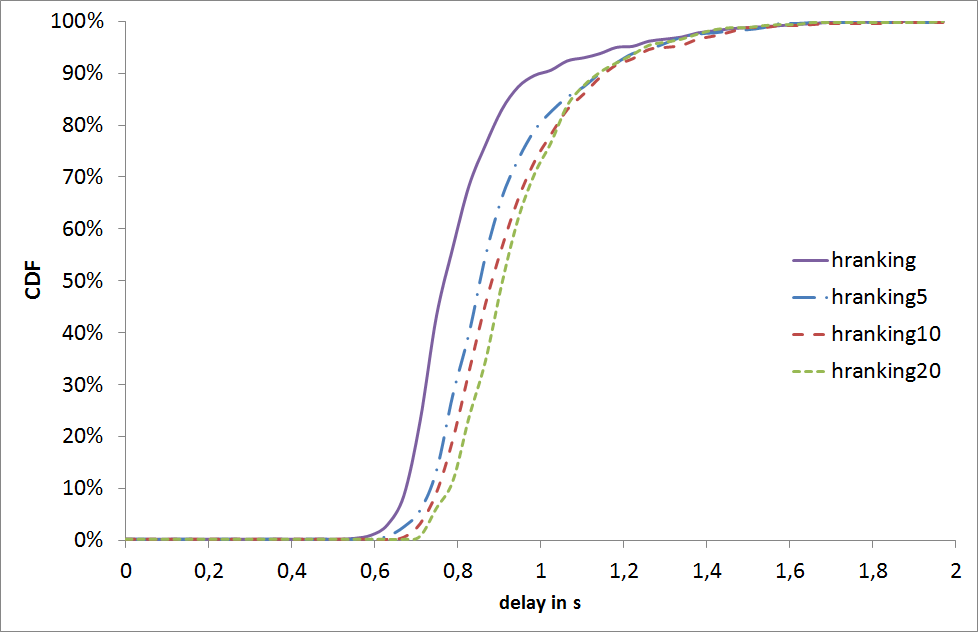}%
\caption{Delay distribution of Hormone Ranking with Hormone clean-up if 5,10, 20 nodes fail}%
\label{fig:delaychurn}%
\end{figure}
\begin{figure}%
\centering
\includegraphics[scale=0.35]{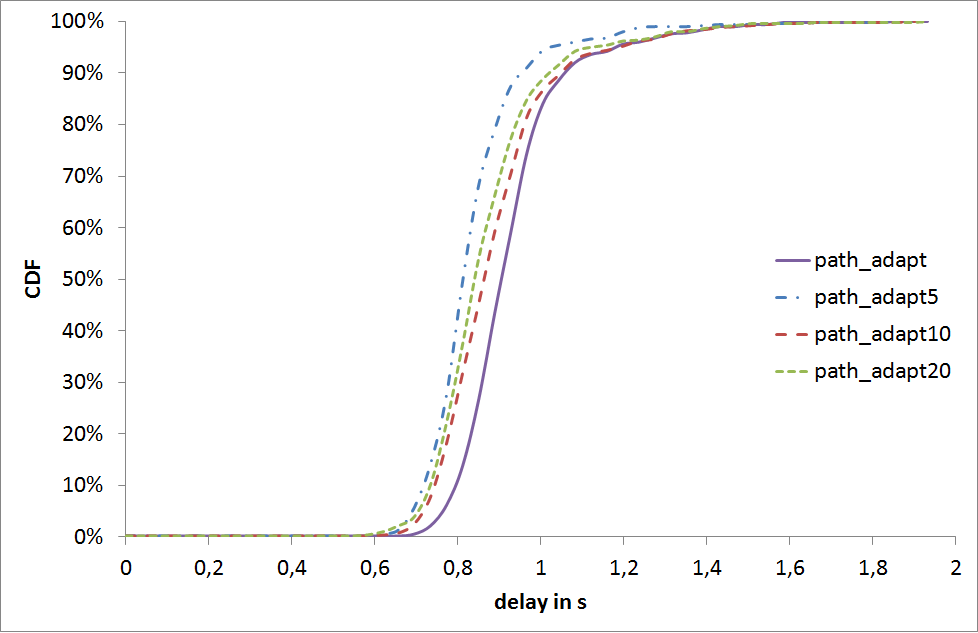}%
\caption{Delay distribution of path adaptive with LRU if 5, 10, 20 nodes fail}%
\label{fig:delaypathchurn}%
\end{figure}
\begin{figure}%
\centering
\includegraphics[scale=0.35]{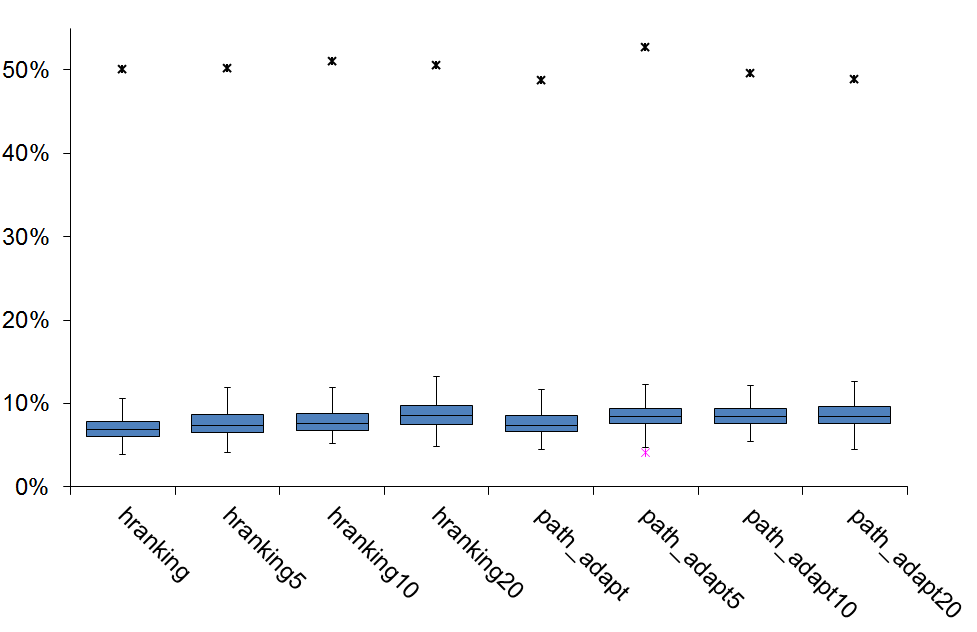}%
\caption{Failed request rate in case of peer churn}%
\label{fig:failedchurn}%
\end{figure}

We simulate churn as nodes being removed regularly one-by-one. We do not handle isolated nodes after peer deletion. Thus, there is a performance gain potential if an overlay algorithm takes care of reconnecting such nodes. We randomly chose 5, 10 and 20 nodes to be removed. \\
We compare hormone ranking with hormone clean-up and path adaptive replication with LRU.\\

Figure~\ref{fig:delaychurn} shows the delay distribution of the hormone ranking algorithm. One can see that the replication algorithm and the delivery algorithm are capable of handling loss. Overall the delay increased a bit, but interestingly the delay for 5, 10 and 20 removed nodes is very similar. However, the failed requests increase slightly in both hormone ranking and path adaptive scenarios, as shown in Figure~\ref{fig:failedchurn}. In general if a keyword is matched by a number of units, a system wide loss of a unit does not have a major impact. \\

The path adaptive algorithm leads to interesting results as shown in Figure~\ref{fig:delaypathchurn}. Here, the delay of the peer churn scenarios is lower than in the original case. This can be explained by referencing the clean-up failures of the original scenario. A clean-up fails if on the current node all units are currently in use or there is no copy of the current unit on a neighbor. A disadvantageous of replica distribution might be that at every second hop a replica is placed, which means that a high number of replicas exist in the system, but because the nodes only see their neighbors, the units get not deleted. If a peer fails, a unit has to move an alternative way. Therefore, the unit movement is increased and more requests can be fulfilled. Path adaptive replication might need an alternative clean-up policy taking a larger neighborhood into account.
\subsection{Scenario 2: 1000 nodes scale-free network}
\begin{figure}%
\centering
\includegraphics[scale=0.35]{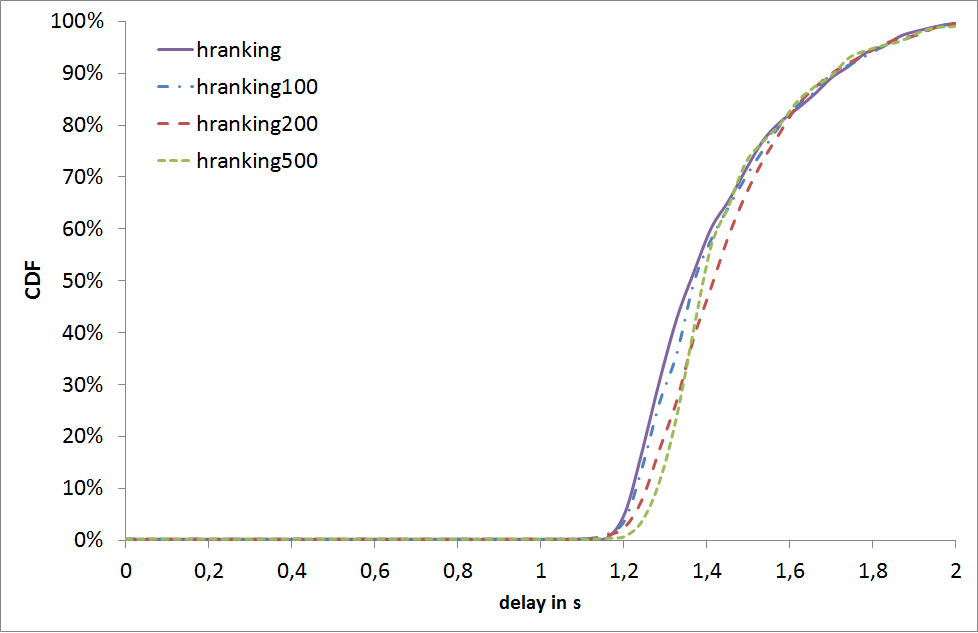}%
\caption{Delay Distribution of Hormone Ranking with Hormone clean-up if 100, 200, 500 nodes fail}%
\label{fig:1000delayhranking}%
\end{figure}
\begin{figure}%
\centering
\includegraphics[scale=0.35]{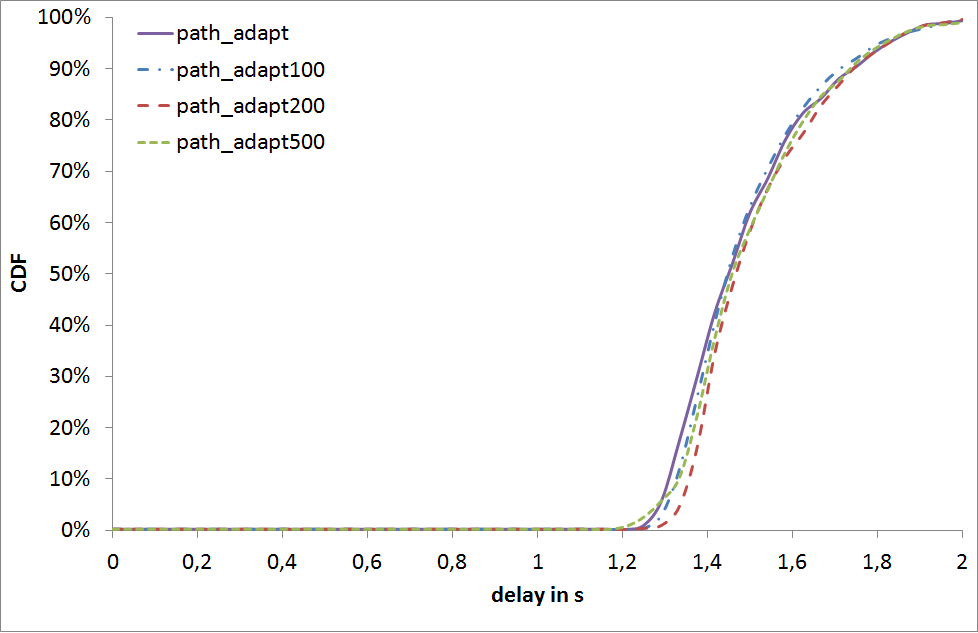}%
\caption{Delay distribution of path adaptive with LRU if 100,200, 500 nodes fail}%
\label{fig:1000delaypathrand}%
\end{figure}
\begin{figure}%
\centering
\includegraphics[scale=0.35]{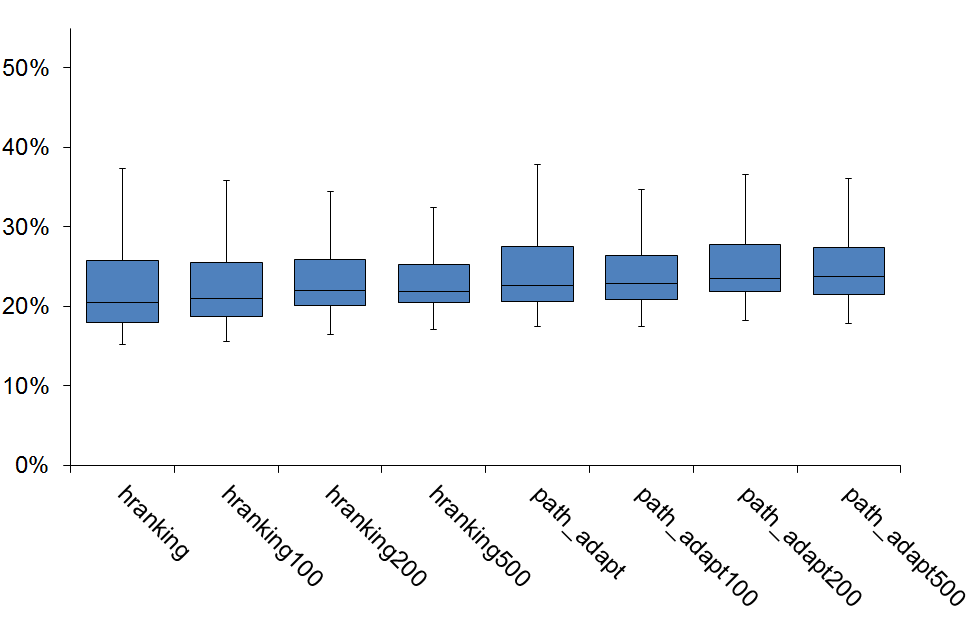}%
\caption{Failed request rate in case of peer churn}%
\label{fig:reqfailure1000}%
\end{figure}
In this section we evaluate the applicability of our delivery algorithm for scale-free networks. We reduce our scenarios to hormone ranking with hormone clean-up and path adaptive replication with LRU. It is shown that the parameters for the 50 node network also work for the 1,000 peer network. Specific optimization using the genetic algorithm could lead to even better results. \\

In Figure~\ref{fig:1000delayhranking} it is depicted that the delay is increased by around 500 ms in comparison to the small network for the hormone ranking algorithm. Furthermore, if 100, 200 and even 500 nodes fail over time the delay does not increase considerably. Note that also high degree nodes may fail, because the nodes leaving the network are chosen randomly. Figure~\ref{fig:1000delaypathrand} shows that the problem with clean-up failures is not experienced as in the 50 nodes network, which can be explained by the network structure and its rather low diameter. Therefore, the delay of path adaptive replication is similar to the hormone ranking algorithm. Both algorithms show slight increases of request failures also in the presence of peer churn (see Figure~\ref{fig:reqfailure1000}). \\

\section{Conclusion}
In this paper we compared different clean-up strategies to be combined with replication mechanisms. We evaluated the delay impact of the clean-up mechanisms, as well as the impact on the utilization of replicas. Furthermore, although reducing the number of replicas, the robustness of the system is still very high. Similar results are reached if the system is applied to a scale-free network of 1,000 nodes.\\

Although the results of the replication mechanisms are promising, the clean-up has a negative impact on the delay, which means that the goals of the clean-up introduction are not reached. Ideally, a combination of replication and clean-up leads to low delay and high utilization. If the settings are as strict as in this paper, path replication with hormone clean-up, although inefficient, performs best. Alternatives could be path adaptive replication with LRU and hormone ranking replication with hormone clean-up. The node failure scenarios showed, that there are still nodes, which block the transport of units. To solve this issue the settings could be less strict regarding the deletion policy. Instead of deleting a unit only if there is a copy of it in the neighborhood, it could be weakened to delete a unit if another unit covering the same keyword is in the neighborhood.

\bibliography{TRSaso}
\bibliographystyle{alpha}
\end{document}